\documentclass[acus]{MOB13}

\usepackage{graphicx}

\begin{document}
\title{THE FLAT BEAM EXPERIMENT AT THE FNAL PHOTOINJECTOR}

\author{D. Edwards, H. Edwards, N. Holtkamp, S. Nagaitsev, 
J.~Santucci, FNAL\thanks{The Fermi National Accelerator Laboratory
is operated under contract with the US Department of Energy}
\\ R. Brinkmann, K. Desler, K. Fl\"{o}ttmann, DESY-Hamburg\\ I.
Bohnet, DESY-Zeuthen, M. Ferrario,  INFN-Frascati}

\maketitle

\begin{abstract}  A technique for production of an electron beam
with a high transverse emittance ratio, a ``flat''
      beam, has been proposed by Brinkmann, Derbenev, and
Fl\"{o}ttmann.\cite{BDF} The cathode of an RF-laser
      gun is immersed in a solenoidal magnetic field; as a result
the beam emitted from a round laser spot
      has a net angular momentum. Subsequent passage through a
matched quadrupole channel that has a 90 degree
      difference in phase advance between the transverse degrees of
freedom results in a flat beam.
      Experimental study is underway at the Fermilab 
Photoinjector. Thus far, transverse emittance ratios as high as 50
have been observed, and the results are in substantial agreement
with simulation.

\end{abstract}

\section{INTRODUCTION}

Two years ago, Ya. Derbenev invented an optics maneuver for
transforming a beam with a high ratio of horizontal to vertical
emittance---a ``flat beam'' ---to one with equal emittances in the
transverse degrees-of-freedom---a ``round beam''.\cite{Derbenev}   
High energy electron cooling at the TeV energy scale was the
motivation.  

Last year, R. Brinkmann and K. Fl\"{o}ttmann of DESY joined with
Derbenev in a paper that reverses the process---obtain a flat beam
from a round beam produced from the cathode of an electron
gun.\cite{BDF}  This could be a significant step toward the
elimination or simplification of the electron damping ring in a
linear collider design.  The other major step in that process is the
delivery of polarized electrons in the flat beam, and this is an
R\&D challenge beyond the scope of the work reported here.

The intent of the present experiment was to demonstrate the
round-to-flat transformation, compare the results with simulation,
and verify that the demonstration was not obscured by other
processes.  In the following sections, we present a simplified
version of the transformation, describe the experimental setup,
present the results, and comment on future plans.

\section{PRINCIPLE}\label{s:princ} Suppose that the cathode of an
electron gun is immersed in a uniform solenoidal field of magnitude
$B_z$.  For the sake of this argument, assume that the thermal
emittance is negligible and ignore RF focusing in the gun.  Then the
particles just stream along the field lines until the end of the
solenoid is reached, at which point the beam acquires an angular
momentum.  A particle with initial transverse coordinates $x_0$,
$y_0$ acquires angular deflections.  With momentum $p_0$ at the
solenoid end, the state of the particle becomes
\[
 \left(
 \begin{array}{c}
  x \\ x' \\ y \\ y'
 \end{array}
 \right)_0
 =
 \left(
 \begin{array}{c}
  x_0 \\ -ky_0 \\ y_0 \\ kx_0
 \end{array}
 \right) 
\label{e:init}\] where
\[ k \equiv \frac{1}{2}\frac{B_z}{(p_0/e)}.
\]

Next pass the beam through an alternating gradient quadrupole
channel.  Assume that the channel is represented by an identity
matrix in the
$x$-direction and has an additional $90^0$ phase advance in $y$.

We get the output state

\begin{eqnarray*}
\left(
\begin{array}{c}
 x \\ x' \\ y  \\ y'
\end{array}
\right) &=&
\left(
\begin{array}{cccc} 1 & 0 & 0 & 0 \\ 0 & 1 & 0 & 0 \\ 0 & 0 & 0 &
\beta \\ 0 & 0 & -\frac{1}{\beta}&0
\end{array}
\right)
\left(
 \begin{array}{c}
  x_0 \\ -ky_0 \\ y_0 \\ kx_0
 \end{array}
 \right) \\ &=&
\left(
\begin{array}{c}
 x_0 \\ -ky_0 \\ k\beta x_0 \\ -\frac{1}{\beta}y_0
\end{array}
\right)\;\;
\stackrel{\beta=1/k}{\rightarrow}\;\;
\left(
\begin{array}{c}
 x_0 \\ -ky_0 \\  x_0 \\ -ky_0
\end{array}
\right).
\end{eqnarray*}

In the last step above, with $\beta = 1/k$, the particles end up
with equal displacements in
$x$ and
$y$ and travelling at equal angles in $x$ and $y$.  This describes a
flat beam inclined at an angle of $45^0$ to the coordinate axes. 
Change to a skew-quadrupole channel, and the flat beam can be
aligned along either the horizontal or vertical axis.

This idealized example is only meant to illustrate the principle. 
The essential points about the quadrupole channel are the $\pi/2$
difference in phase advance between the transverse
degrees-of-freedom, and the match of the Courant-Snyder
parameters.\cite{burov}  This may be accomplished with as few as
three quadrupoles.  Of course, in practice, RF focusing fields in
the gun and in a booster cavity, space charge, and so on cannot be
ignored.

\begin{figure*}[t]
\centering
\includegraphics*[width=120mm]{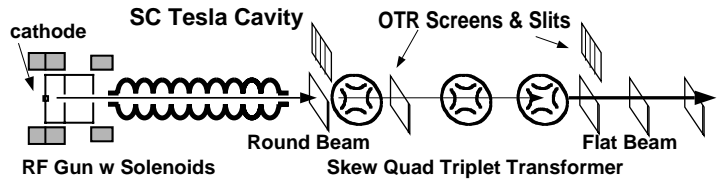}
\vspace{-5.0in}
\caption{Very schematic rendition of the layout at Fermilab related
to this experiment.}
\label{f:a0}
\end{figure*}

With the inclusion of thermal emittance, Brinkmann, Derbenev, and
Fl\"{o}ttmann\cite{BDF} speak of an achievable emittance ratio of
order $10^2$ or more for a beam with normalized emittance
$\sqrt{\epsilon_x\cdot\epsilon_y}\approx 1 \mu$m per nC of bunch
charge.  The expression for the emittance ratio is
\[
     \frac{\epsilon_x}{\epsilon_y}\approx\frac{4k^2\sigma_c^2}
                                        {\sigma_c'^2}
\]
where now in the definition of $k$, $B_z$ remains the field on
the cathode, but $p_0$ is the momentum at entry to the quadrupole
channel, and $\sigma_c$, $\sigma_c'$ are the standard deviations of
the distribution in displacement and angle at the cathode.  
The resulting
vertical emittance would be 0.1
$\mu$m, in the range of interest for a linear collider.  Liouville's
Theorem remains in effect for the 4-dimensional transverse
emittance, but the angular momentum provides the lever by which
emittance may be moved from one degree-of-freedom to another.

\section{THE FERMILAB PHOTOINJECTOR ENVIRONMENT}

The photoinjector at Fermilab is well suited to this sort of
experiment. 
 The RF gun delivers electrons with a kinetic energy of (typically)
3.8 MeV.  The superconducting booster cavity  raises the electron
energy to 17 MeV.

The solenoid is composed of three separately excited coils permitting
fields at the cathode in the range 0 to 2.7~kG.    The coil
immediately upstream of the cathode, the ``bucker'', is normally
excited with current opposite to that of the next coil, the
``primary'' to produce zero field at the cathode.  Downstream, the
combination yields solenoidal focusing, which can be adjusted with
the third coil, the ``secondary''.  The secondary has little effect
on the field on the cathode.

Following the booster cavity, about 8 meters of beamline are
available for experiments. There are 11 quadrupoles that are easily
moved about or rotated into the skew orientation.  A dozen view
screens are situated on the line, and there are three locations
where slits are installed for emittance measurement.  
The laser can operate at a variety of pulse lengths up to 12~ps, the
setting that we used.  Bunch charge as high as 10 nC is available. 
We operated at no higher than 1 nC in order to reduce space charge
effects as much as possible. 
The layout as related to this experiment is sketched in
Fig.~\ref{f:a0}.

\section{PROCEDURE}

The solenoid coils were set to produce a field at the cathode in the
expected range, about 0.75 kG.  Using the language of the preceding
section, this meant setting the bucker to zero current and
controlling the cathode field with the primary.  The beam was
observed at the location of the two screens immediately downstream
of the booster cavity, and by adjustment of the secondary coil, the
beam spot was made the same size at these two places.  In other
words, a beam waist was produced.  At this stage, the beam has a
round shape on the screens.

The simple argument of Sec.~\ref{s:princ} is no longer valid for
determination of the $\beta$ for the match, because the solenoid
field is not uniform and the RF focusing and acceleration must be
taken into account.  Making use of linearity, axial symmetry, and the
conservation of canonical angular momentum between the cathode and
the waist yields for the value of $\beta$ at entry to the quadrupole
channel
\[
    \beta = \frac{\sigma_w^2}{\sigma_c^2}\frac{2 (p_w/e)}{B_c}
\] where the subscripts $c$ and $w$ refer to the cathode and waist
respectively and the $\sigma$'s characterize the radii of the beam
spots.  The other Courant-Snyder parameter involved in the match,
$\alpha$, is zero due to the choice of a waist as the match point.

Given preliminary values for the matching parameters, an
(asymmetric) skew triplet was set up.  Flat beam
profiles were rather easily achieved by adjustment of available
tuning parameters, including the launch phase from the RF gun. The
latter proved to be particularly important, a circumstance that is
yet to be explained.  

\section{RESULTS}

\begin{figure}[htb]
\centering
\includegraphics[height=30mm]{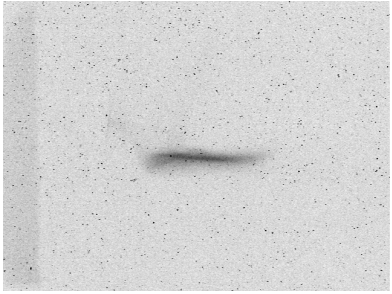}
\vspace{0.4in}
\caption{Beam profile on OTR screen 1.2 m downstream of the third
skew quadrupole.}
\label{f:otr1.2}
\end{figure}

\begin{figure}[htb]
\centering
\includegraphics[height=30mm]{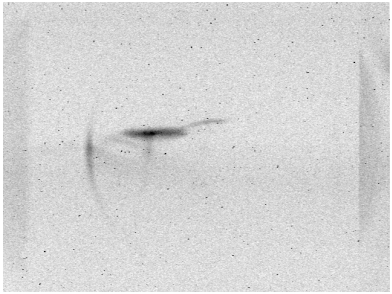}
\vspace{0.4in}
\caption{Beam profile on OTR screen 3.6 m downstream of the third
skew quadrupole.  Dark current is visible to the right of the main
beam image.}
\label{f:otr3.6}
\end{figure}

The transformation should work --- it's linear dynamics --- and it
does. The match and phase difference were achieved with three skew
quadrupoles.  The beam image on an OTR screen 1.2~m downstream of
the third quadrupole is shown in Fig.~\ref{f:otr1.2}; the beam width
is an order of magnitude larger than the height.  A critical
observation is that the beam remain flat as it drifts farther
downstream.  That it does is demonstrated in Fig.~\ref{f:otr3.6}
near the end of the beamline at 3.6~m from the third quadrupole.


In Fig.~\ref{f:otr1.2} there is a hint of an s-shape, which likely
indicates that spherical aberrations (e.g. space charge) are at
work.  If the solenoid field on the cathode is varied up or down
from the matched condition the beam apparently rotates clockwise or
counterclockwise as it drifts, indicating that the angular momentum
is no longer completely cancelled.  Of course, it isn't a real
rotation --- there's no torque--- it's a shear.

\begin{figure}[htb]
\centering
\includegraphics[width=71mm]{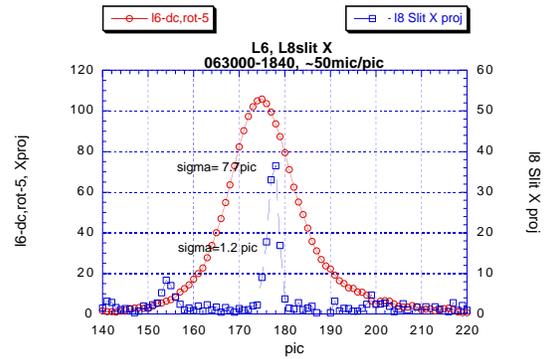}
\caption{Projection of images used in emittance measurement at slit
location and downstream of slit system.}
\label{f:res}
\end{figure}

In these figures, the beam is flat in the horizontal plane.  The OTR
screens are viewed from the side, and so a beam that is flat
horizontally presents a depth of field problem for best emittance
analysis.  So in later stages of the experiment, the beam was made
flat in the vertical plane. From slit data in this orientation, the
measured ratio of emittances is about 50:
$\epsilon_x\approx 0.9\mu$m, $\epsilon_y \approx 45\mu$m, with the
one degree-of-freedom normalized emittance defined by
$\epsilon^2=\gamma^2 (v/c)^2 (\langle x^2\rangle\langle x'^2\rangle
-\langle x x'\rangle^2)$.  We feel that this is a good result for an
initial experiment.  The horizontal emittance measurement is
resolution limited, as illustrated in Fig.~\ref{f:res} wherein a
sequence of slit images is superimposed in order to form a
distribution.  The standard deviation of the narrow distribution is
comparable to a single pixel of the CCD camera viewing the screen.

The product of the emittances is higher than that usual in 
operation with round beams;  typically, the emittance in each
transverse degree-of-freedom is about 3 to 4 $\mu$m.  However, there
is no reason to believe that the emittance compensation normally in
use would be effective under the conditions of this experiment.

The simulations\cite{Astra},\cite{Sergei} carried out prior to the
measurements provided useful guidance, but were not perfect.  The
prediction of spot size just downstream of the gun worked fine.  
But to achieve the match to the quadrupoles, the solenoid required
adjustment.

In order to obtain agreement between the location of the beam waist
downstream of the booster cavity, a modification of the focusing
characteristics of this device was required.  In the Chambers
approximation\cite{Chambers}, its demagnification is a factor of 5,
so its treatment is sensitive to a number of factors, e.g. the exact
field
profile.  It will be
worthwhile to measure the transfer matrix through the cavity
experimentally.

\section{CONCLUSIONS} The round-to-flat transformation has been
verified, with a demonstrated emittance ratio of a factor of 50
between the two transverse degrees-of-freedom.  Further work will be
needed to restore the emittance compensation necessary to the
delivery of low transverse emittance, and that is the subject of a
follow-on experiment, in the direction suggested by Brinkmann,
Debenev and Fl\"{o}ttmann in their EPAC2000 paper.\cite{BDF2}  The
predictive capability of the simulations is encouraging thus far,
and the results reported here indicate directions for improvement.

\section{ACKNOWLEGEMENTS} Support of the Fermilab and DESY
managements is gratefully acknowledged.  Thanks to  Jean-Paul
Carneiro, Mark Champion, Michael Fitch, Joel Fuerst and Walter
Hartung for their invaluable help in the operation.

\end{document}